# Stochastic reconstruction of protein structures from effective connectivity profiles

Katrin Wolff[1], Michele Vendruscolo[2] and Markus Porto*[1]

Address: [1]Institut für Festkörperphysik, Technische Universität Darmstadt, Hochschulstraße 6, 64289 Darmstadt, Germany and [2]Department of Chemistry, University of Cambridge, Lensfield Road, Cambridge CB2 1EW, UK

Email: Katrin Wolff - wolff@fkp.tu-darmstadt.de; Michele Vendruscolo - mv245@cam.ac.uk; Markus Porto* - porto@fkp.tu-darmstadt.de

* Corresponding author

## Abstract

We discuss a stochastic approach for reconstructing the native structures of proteins from the knowledge of the "effective connectivity", which is a one-dimensional structural profile constructed as a linear combination of the eigenvectors of the contact map of the target structure. The structural profile is used to bias a search of the conformational space towards the target structure in a Monte Carlo scheme operating on a $C\alpha$-chain of uniform, finite thickness. Structure information thus enters the folding dynamics via the effective connectivity, but the interaction is not restricted to pairs of amino acids that form native contacts, resulting in a free energy landscape which does not rely on the assumption of minimal frustration. Moreover, effective connectivity vectors can be predicted more readily from the amino acid sequence of proteins than the corresponding contact maps, thus suggesting that the stochastic protocol presented here could be effectively combined with other current methods for predicting native structures. PACS codes: 87.14.Ee.

## Introduction

The challenges presented by the protein folding problem have been approached from a wide range of different theoretical angles. Computationally convenient structural representations of protein structures such as lattice models [1,2] or Gō-models [3,4] have the advantage of simplicity and of capturing some of the more universal properties of the protein folding process. However, lattice models severely restrict the conformation space, and Gō-models consider primarily interactions between pairs of amino acids that are in contact in the native structure. This latter approach is inspired by the "minimal frustration" view [5], in which the assumption is made that non-native interactions play a relatively minor role in shaping the free energy landscape of pro-teins. As an alternative, sophisticated force fields that employ all atom representations of protein structures [6] provide results in good agreement with a range of experimental observations, but are computationally demanding and thus tend to be restricted to the study of the faster among the dynamical processes experienced by proteins.

In order to investigate whether an alternative trade-off between simplicity and accuracy can be found, in the present work we adopt an approach based on the use of structural profiles [7-11]. In this work, we begin to study this problem by considering the reconstruction problem from exact structural profiles, prerequisite to the investigation of folding energy landscapes.

Reconstruction as discussed in this study always operates on an explicit structure description and thus a successful "reconstruction" is equivalent to a folding process reaching the native structure. In this sense, our approach is similar to G̅n-models as it uses information derived from the folded structure but dissimilar in the important fact that interactions are non-specific, i.e. there are interactions between all residues, not only native ones. As it is the case of G̅n-models, an agreement between the dynamics of folding found experimentally and computationally would imply that the native structure of a protein largely dictates the path of folding [5,12].

Another important reason for exploring the use of structural profiles is the correlation to (optimal) sequence hydrophobicity [10] allowing prediction of profile from sequence alone (without prior prediction of contact maps) with relatively high accuracy [13,14]. These results open the possibility of using predicted structural profiles to aid, in turn, the prediction of the three-dimensional coordinates of the native states of proteins. One of the major problems in this type of approach, however, is that structural profiles can be predicted with only limited accuracy. Therefore it is key to establish procedures which are in principle able to determine three-dimensional conformations from the knowledge of approximate, or noisy, structural profiles. Stochastic methods are the most promising ones for solving this problem. In this work we make a step in establishing this approach by studying the problem of reconstruction of three-dimensional structures from structural profiles using a Monte Carlo procedure for a set of representative small proteins. A major advantage over work that we previously carried out [15] is achieved by using a more general structure profile and by restricting the structural profile to amino acids that form cooperative contacts and, in this sense, show protein-like behaviour, resulting in a more reliable reconstruction.

Even though our method is not meant to perform structure predictions on its own, the results presented here suggest that it might be convenient to incorporate predictions of the effective connectivity vectors as additional information into methods for predicting the native structures of proteins, in particular those based on the molecular fragment replacement procedure [16-18], which are currently the most effective ones for achieving this goal [19].

**Contact maps and effective connectivity**

The contact map corresponding to a given structure of a protein is a $N \times N$ binary symmetric matrix, where $N$ is the number of amino acids, storing information about the amino acid pairs that are in contact. The matrix entry $C_{ij}$ is set to 1 if amino acids $i$ and $j$ are in contact and 0 otherwise. Two amino acids are defined as being in contact if the distance $x_{ij} = |\vec{r}_{ij}|$ between their $C_\alpha$-atoms is less than a threshold value of $r_c = 8.5\text{Å}$,

$$C_{ij} = \begin{cases} 1 & x_{ij} < r_c \\ 0 & x_{ij} \geq r_c \end{cases}. \qquad (1)$$

The structural profile employed here is the effective connectivity (EC) $\mathbf{c}$ [11], which is a linear superposition of the eigenvectors of the contact map, $\mathbf{v}^{(j)}$, weighted by a function of the respective eigenvalues $\lambda^{(j)}$,

$$c_i = \frac{1}{A} \sum_{j=1}^{N} \frac{1}{\Lambda - \lambda^{(j)}} v_i^{(j)} \langle v^{(j)} \rangle, \qquad (2)$$

where $\langle v^{(j)} \rangle$ denotes the average of vector $\mathbf{v}^{(j)}$ and $A$ is a normalising constant,

$$A = \sum_{j=1}^{N} \frac{1}{\Lambda - \lambda^{(j)}} \langle v^{(j)} \rangle^2, \qquad (3)$$

such that $\langle c \rangle = 1$. The parameter $\Lambda$ is determined such that the relative variance of the EC equals the relative variance of the contact vector $\tilde{\mathbf{c}}$ [11],

$$\frac{\langle c^2 \rangle}{\langle c \rangle^2} = \frac{\langle \tilde{c}^2 \rangle}{\langle \tilde{c} \rangle^2}, \qquad (4)$$

the latter being defined as the number of contacts of amino acid $i$,

$$\tilde{c}_i = \sum_{j=1}^{N} C_{ij}. \qquad (5)$$

The largest contribution to the EC comes from the principal eigenvector (PE), the eigenvector $\mathbf{v}^{(1)}$ to the largest eigenvalue $\lambda^{(1)}$, and for single domain proteins the correlation between PE and EC is very high [11]. Since during the folding process single-domain proteins populate non-com-pact conformations, the use of the more general EC is preferable over that of the PE. For the small proteins investigated in this work, this distinction is of minor importance as folding is likely to happen as a single cooperative event. When, in the future, progressing to larger proteins, folding may start independently at different sites and we expect the more general EC to fare better. Both structural profiles (PE and EC) contain information about the connectivity of each amino acid. Well connected residues tend to have larger entries in the structural profile than those connected to fewer residues. This fact also explains the correlation to hydrophobicity, as residues with many contacts are buried inside the protein fold (see discussion in [10]).

### Restriction to cooperative contacts

Protein structures can be identified by their respective structural profile. For the PE, this matching has been shown to be unique for a set of representative proteins up to 120 residues in length [9] and we expect the EC, too, to uniquely determine the structure. In fact, the EC has been successfully used to perform structural alignments of proteins [20]. In a stochastic reconstruction approach, however, the ruggedness of the landscape corresponding to a cost function based on the difference of structural profiles may pose severe difficulties. In previous work we found that compact structures without discernible secondary structure often resulted in dead ends in Monte Carlo simulations. Since the EC is more suitable for describing folded structures, we restrict the computation of the EC at any time step in the simulation to those parts of the structure that exhibit cooperative contacts (definition see below) characteristic of secondary structure elements. The target EC is computed with the same restrictions but with respect to the target structure.

The typical contact pattern of $\alpha$-helices consists of successive contacts between amino acids $i$ and $i + A$ with $A = 3$ or $4$. For the assignment of the existence of an $\alpha$-helix we require at least four such successive contacts with a contact threshold of 8.5Å. Additionally, positive chirality is requested, $r_{i-1,i} \cdot (r_{i,i+1} \times r_{i+1,i+2}) > 0$. Existence of $\beta$-sheets for individual residues is characterised by detecting contact patterns of $i$ and $i + A$ ($A \geq 5$, parallel $\beta$-sheets) or $i$ and $B$ - $i$ ($B \geq 7$, anti-parallel $\beta$-sheets) with at least four consecutive contacts and fixed $A$, $B$. To make up for the additional condition of chirality for helices, the contact threshold for $\beta$ sheets is set to 7Å.

The contact map is restricted to those residues that show cooperative contacts by deleting all rows and columns without such contacts. In the EC the corresponding entries are similarly set to zero (see Figs. 1 and 2) and also ignored in the computation of the averages $\langle v^{(j)} \rangle$. As amino acids that do not show the cooperative behaviour described above are disregarded, their contacts in a successful reconstruction are also not constrained and their positions restricted only by the requirement of maintaining the chain connectivity (and angles adopting allowed values). For single non-cooperative amino acids in the protein core these conditions are sufficient to provide relatively high resolution. To reduce the number of such non-cooperative amino acids an additional "secondary structure" assignment is introduced for regions in the proximity of $a$-helices that have the same contact pattern but lack correct chirality. This assignment method, when com-pared to DSSP [21] or STRIDE [22] (as implemented in VMD [23], see Fig. 1), tends to slightly overestimate the size of secondary structure elements. This effect, at least in the present context, is convenient for the reasons just mentioned. It should also be emphasised that the main objective of these admittedly *ad hoc* definitions lies not in the accurate assignment of secondary structure to a chain of $C_\alpha$-atoms but in the capture of characteristic features of protein folds to permit the effective calculation and comparison of ECs. The actual information about specific secondary structure (i.e. whether $\alpha$-helix or $\beta$-sheet) from the target is not used in the reconstruction respectively folding dynamics.

### Energy function for Monte Carlo simulations

In order to represent protein structures in a computationally effective way, we used here a simplified version of the tube model [24]. This model consists of a chain of $C_\alpha$-atoms (distance 3.8Å between consecutive $C_\alpha$-atoms) within a tube of uniform thickness (diameter 3.3Å) No amino acid specific properties enter the model. The reconstruction is thus driven by the energy provided by the structural profile,

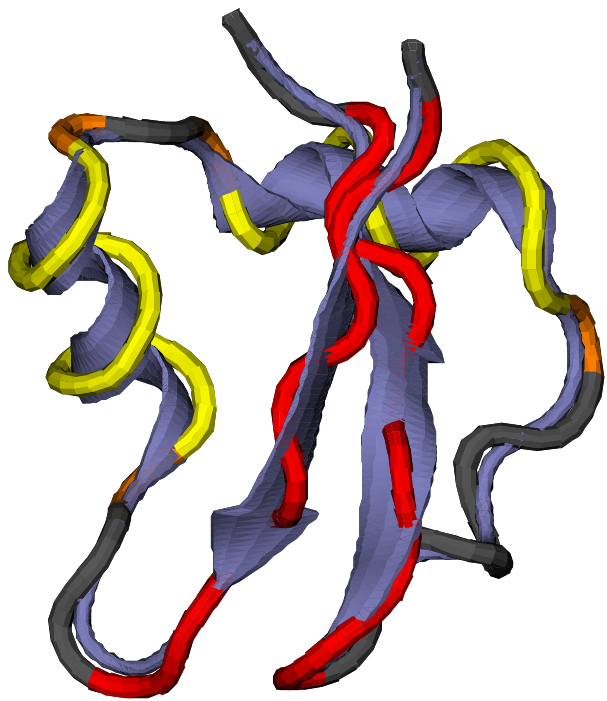

**Figure 1**
**Cooperative contacts in 1e0g**. Comparison between the cooperative contacts in the LYSM domain from *E. coli* MLTD (PDB code 1e0g) and the secondary structure assignment by STRIDE [22] (as implemented in VMD [23]). Red: $\beta$-sheet, yellow: $\alpha$-helix, orange: regions in proximity of secondary structure elements, grey: other residues.

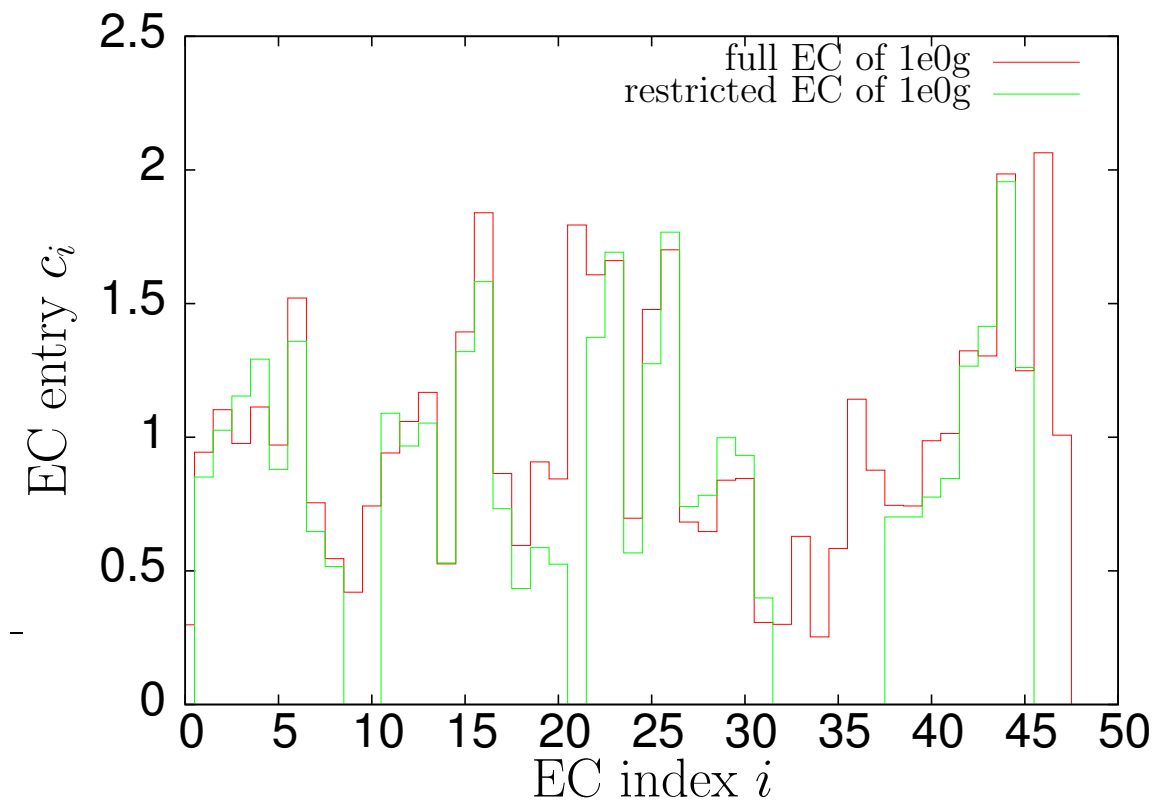

**Figure 2**
**Comparison between full and restricted EC profiles**. Comparison between full and restricted EC profiles for a representative target structure (PDB code 1e0g). Although some of the large entries of the effective connectivity vanish the remaining entries change only moderately.

$$E_{\mathrm{EC}} = \omega \sum_{i=1}^{N} |c_i - t_i|, \qquad (6)$$

where we compare the restricted EC **c** of the conformation in the current Monte Carlo step to the restricted EC **t** obtained from the structure to be reconstructed, and define the energy as the sum of differences in the vector entries. The factor is set to 10 such that $E_{\mathrm{EC}}$ is of the order of 1 per residue for non-native structures. The energy $E_{\mathrm{EC}}$ favours the formation of contacts and leads to a contraction of the chain. The tube model therefore can be considerably simplified to only account for steric exclusion and bending rigidity. We do not consider hydrophobic energy and geometric constraints for hydrogen bonds. The steric energy $E_{\mathrm{steric}}$ is reduced to a term prohibiting too tight turns and tube overlaps [24],

$$E_{steric} \begin{cases} \infty & r_{i,i+1,i+2} < 2.5 \ ^- \ \text{ or tube overlap} \\ 0 & \text{otherwise} \end{cases}, \quad (7)$$

where $r_{i,i+1,i+2}$ is the radius of the circle defined by residues $i$, $i + 1$ and $i + 2$.

The restriction of the EC to residues that show cooperative contacts partly takes the role of the hydrogen bond term by giving rise to $\alpha$-helices or $\beta$-sheet secondary structure elements. We found that $\alpha$-helices in this model tend to be deformed and therefore introduced another energy term $E_{helix}$ that favours more regular $\alpha$-helices by penalising helix-residues (i.e. residues that show cooperative contacts reminiscent of $\alpha$-helices) whose contacts were more distant than 6.5Å (though below 8.5Å) and rewarding those that were closer. The helix energy $E_{helix}$ is defined as

$$E_{helix} = 0.5N_H - N_{H2}. \quad (8)$$

Here, $N_H$ is the number of helix residues and $N_{H2}$ the number of helix residues in close contact. By doing so, helices on average get no energy penalty nor reward and their ratio to $\beta$-sheets is not shifted.

The total energy $E_{tot}$ used in the Metropolis Monte Carlo simulations consists of the steric energy term $E_{steric}$, the helix term $E_{helix}$ and the EC term $E_{EC}$,

$$E_{tot} = E_{steric} + E_{helix} + E_{EC}. \quad (9)$$

Temperature was between 0.7 and 0.9 in units of the above energy scale and kept fixed during folding simulations.

### Structure set

In order to assess the efficiency of the reconstruction procedure we compiled a representative set of small proteins from all Protein Data Bank (PDB) protein entries of length between 25 and 50 amino acids. Structures that are not readily represented by the version of the tube model that we adopt here (e.g. because of atypical $C_\alpha$-distances, bond angles or tube overlaps) are discarded, as are proteins with less than 70% of amino acids showing cooperativity. These latter proteins are the most common case among discarded structures. This selection results in 1507 proteins out of the possible 3706 small ones in the PDB. These 1507 proteins are grouped according to their SCOP [25] folds, and for SCOP classes *a*, *b*, *d* and *g* (all-$\alpha$, all-$\beta$, $\alpha + \beta$ and "small") the longest representative of each fold is chosen as reconstruction candidate. SCOP class *c* ($\alpha/\beta$) did not appear among the proteins, the other classes were ignored. This leaves us with 54 representative protein structures of lengths between 31 and 50 amino acids (for distribution of lengths and classes see Figs. 3 and 4).

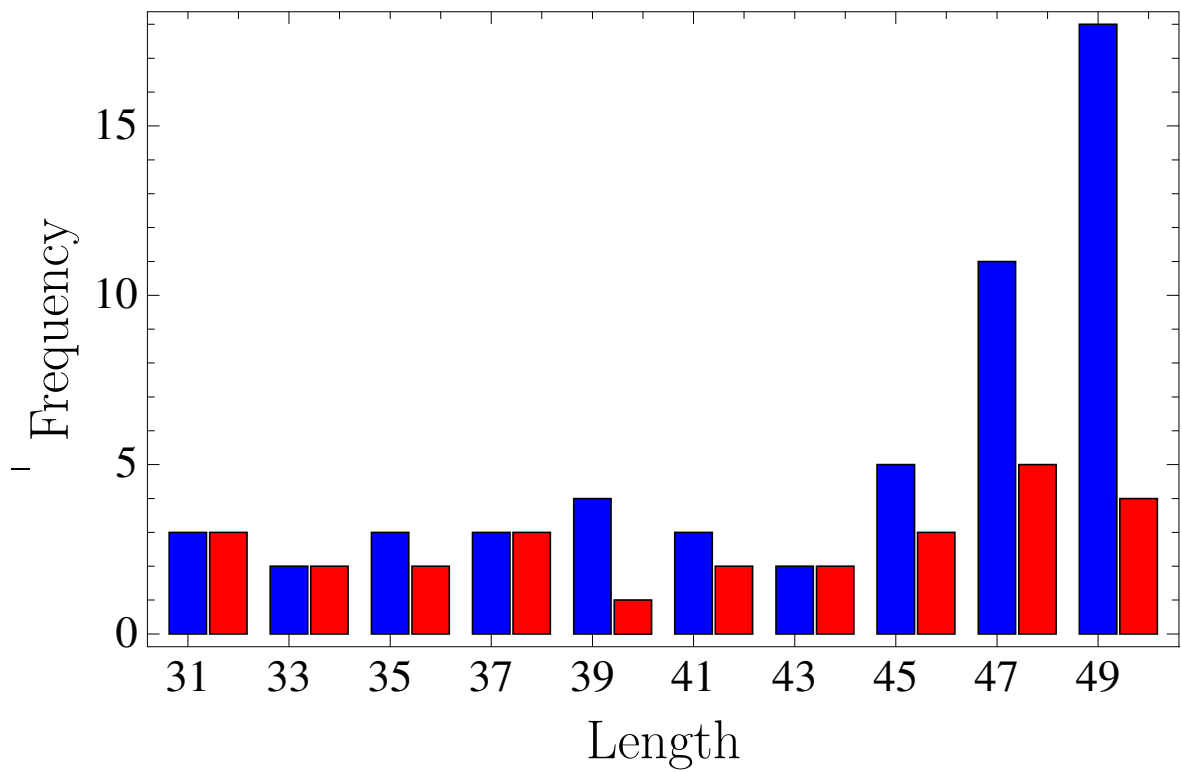

**Figure 3**
**Length distribution of target proteins**. Histograms of the lengths of the target proteins (blue) and of the number of the corresponding reconstructed proteins (red).

## Results and discussion

### *Analysis of successful reconstructions*

A reconstruction attempt was considered to be successful if a structure of the same restricted EC as the target structure had been found and therefore $E_{EC} = 0$. With the choice of $\omega = 10$ in Eq. (6) this definition is equivalent to the minimum of energies $E_{tot}$ encountered up to that point of the simulation. For all structures observed this also meant that the restricted contact map had been recovered completely (contact overlap $q = 1$),

$$q = \frac{|Q_t \cap Q_c|}{\max\left(|Q_t|, |Q_c|\right)}. \qquad (10)$$

Here $Q_t$ is the set of contacts in the target and $Q_c$ the contacts in the present structure. This corresponds to a contact overlap $q_{full}$ to the full (unrestricted) contact map of 74% to 100% for the various proteins (see Fig. 5). This percentage of recovered native contacts can be enough to allow reconstruction to good resolution [26].

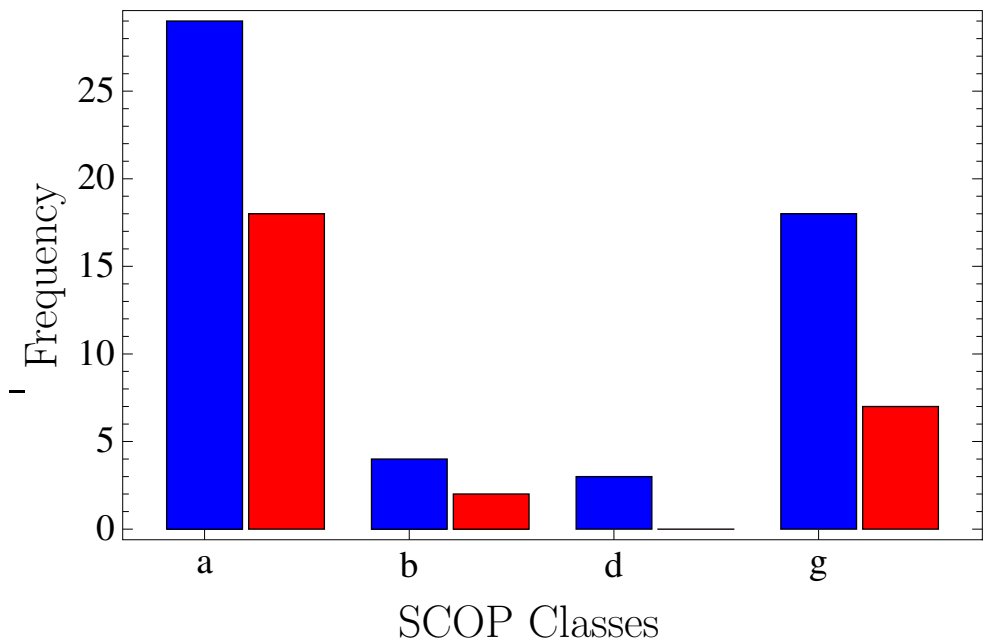

**Figure 4**
**Class distribution of target proteins**. Histograms of the number (blue) of target proteins in SCOP classes *a* (all *α*), *b* (all-*β*), *d* (*α* + *β*) and *g* (small), and of the number (red) of the corresponding reconstructed proteins.

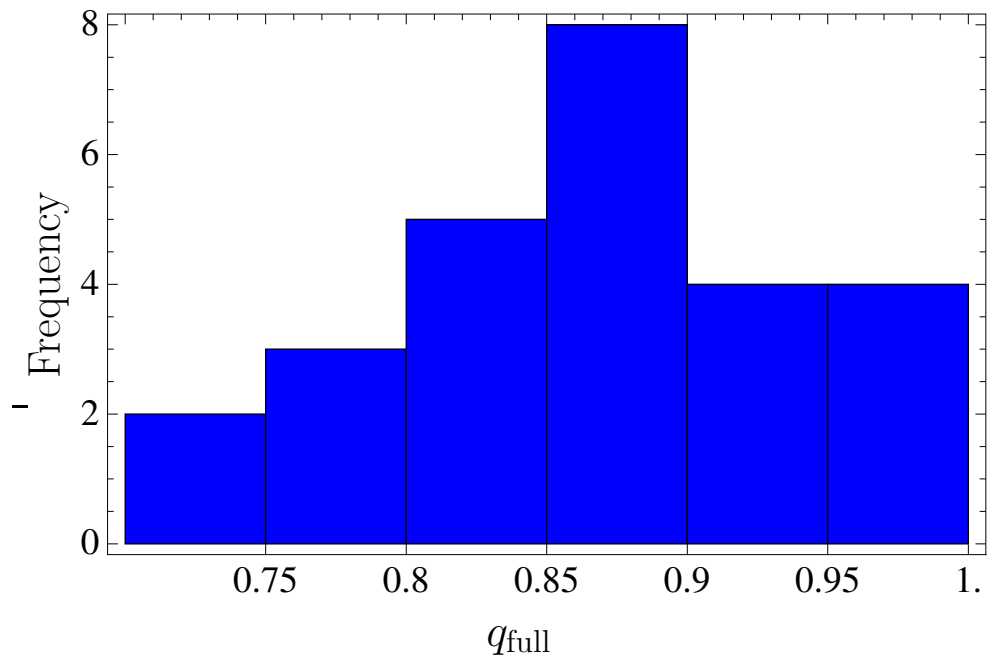

**Figure 5**
**Distribution of contact overlap**. Distribution of full contact overlaps, i.e. not restricted to cooperative contacts, for the cases of successful reconstructions discussed in this work.

The root mean square deviation (RMSD) of $C_\alpha$ atoms ranges between 2.2Å and 9.6Å over the entire chain. We explain these variations by the following considerations. The omission of unstructured tails at the ends improves the overall RMSD to between 1.5Å and 9.1Å. More importantly, there were many non-compact structures in the structure set. Compactness appears to be the crucial condition for low RMSDs, even more than a high content of secondary structure. Thus, compact structures (Fig. 6) were reconstructed to low RMSDs while less compact ones (Fig. 7) were considerably deformed even though $q_{full}$ > 90%. This behaviour can often be observed for single $\alpha$-helices or two-bundles as the helices that are at most weakly inter-connected can be deformed considerably without breaking the contact pattern (see Fig. 7). Here, the resolution is almost as good as a Gn-model of the full, unrestricted structure would achieve. In the case of $\alpha$-helices the problem could likely be eased by an improved version of the helix energy $E_{helix}$, for example by enforcing straight helix axes, but non-compact structures were generally problematic. Other structures can nicely be aligned piecewise to their targets but the relative orientations of the parts are poor resulting in a misleadingly high RMSD. Despite the sometimes large RMSD, no true ambiguity was found in the structure set, i.e. no alternative structures of the same EC.

### Reconstruction statistics

Successful reconstruction is not distributed evenly among the four SCOP classes covered by the candidate set (Fig. 4). All-$\alpha$ proteins were most abundant in the set and also easiest to reconstruct (18 out of 29). Out of only four all-$\beta$ proteins two could be reconstructed while none of the three $\alpha$ + $\beta$-proteins and six out of 18 small proteins (SCOP class $g$) were successful. Of these six proteins two had only $a$ secondary structure, two only $\beta$ and two both. The secondary structure assignment did not by itself favour $\alpha$-helical over $\beta$-sheet contacts, but it was observed that once a secondary structure element had been formed it remained relatively stable for the rest of the simulation. This situation turned out to be problematic for the formation of cooperative contacts between amino acids distant along the chain, in particular in the case of $\beta$-sheet contacts between distant strands, while it was irrelevant for more local patterns like $\alpha$-helices or $\beta$-hairpins. Only for very small structures (see Fig. 8) $\beta$-sheets could be recovered. Increasing the length also expectedly made reconstruction more difficult owing to the larger conformation space to be sampled (Fig. 3) – and also slowed down computation due to larger eigensystems to be solved. In total, 26 of 54 protein structures were reconstructed from their respective EC profiles.

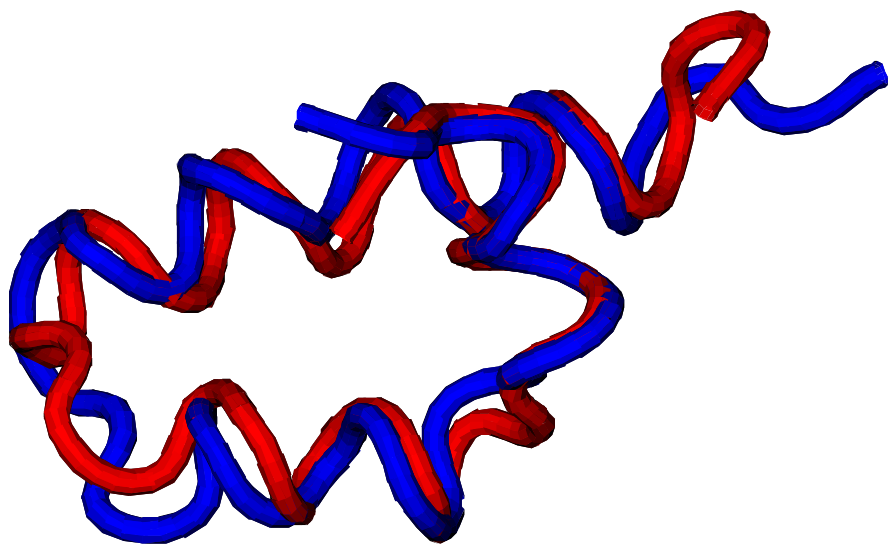

**Figure 6**
**Reconstruction of 1klv**. Comparison between the target (red) and reconstructed (blue) structures of the DNA-binding domain of MafG (PDB code 1klv). The RMSD between the two structures is 2.4Å.

## Conclusion

We have presented a stochastic scheme to reconstruct the three-dimensional structures of proteins from the knowledge of their effective connectivity vector. We have demonstrated that in its current implementation this method is rather effective for proteins in the all-$\alpha$ fold class but shows limitations for more complex proteins with non-local cooperative contacts. Since the stochastic method employed here was based on Monte Carlo simulations at fixed temperature, more advanced sampling techniques are expected to provide improved results, particularly for longer proteins. In addition, further improvements might be achieved by enhancing the sampling of long-range cooperative contacts.

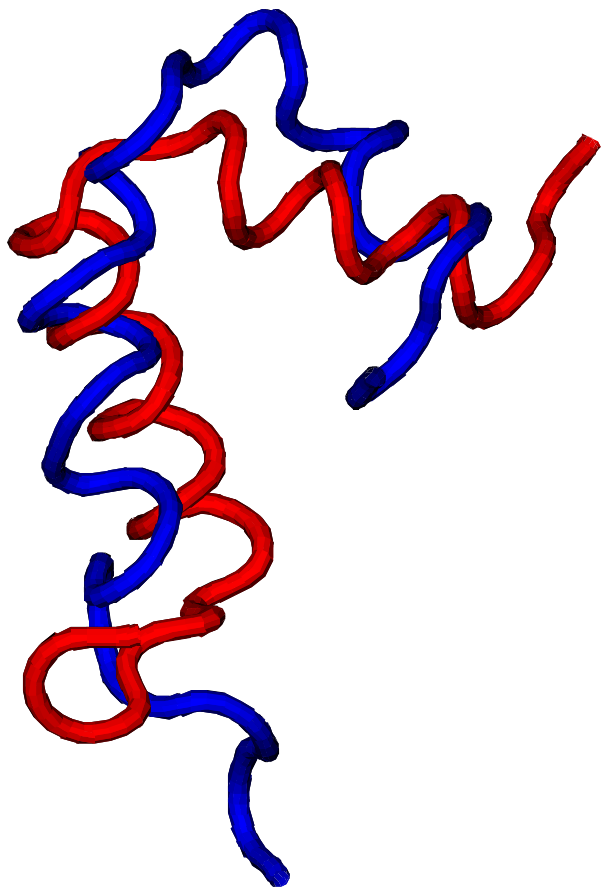

**Figure 7**
**Reconstruction of 1rqt, chain B**. Comparison between the target (red) and reconstructed (blue) structures of chain B of the dimeric N-terminal domain of ribosomal protein L7 from *E. coli* (PDB code 1rqt, chain B). The RMSD between the two structures is 7.0Å.

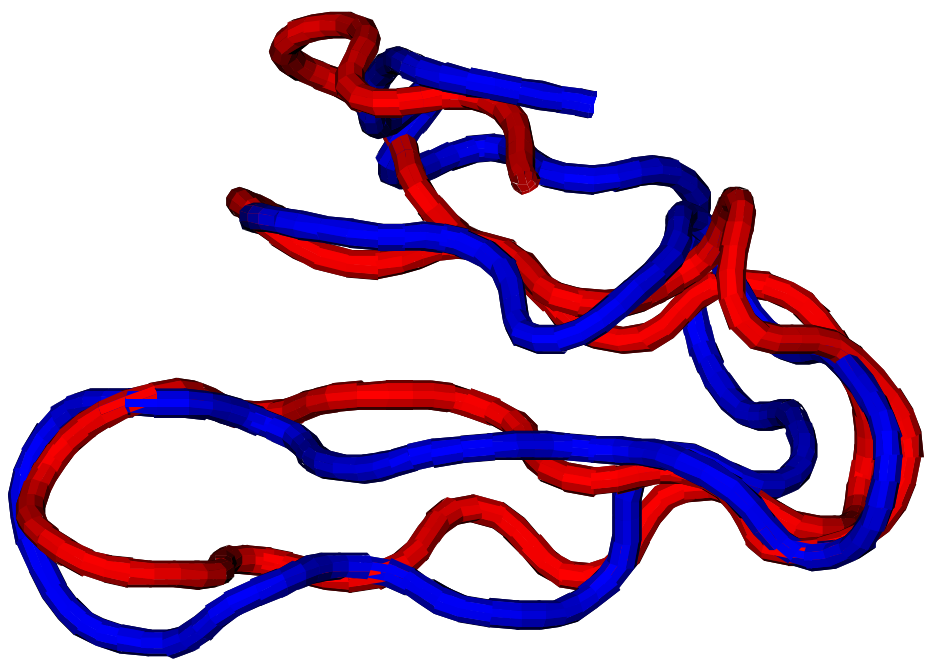

**Figure 8**
**Reconstruction of 2i2v, chain 4**. Comparison between the target (red) and reconstructed (blue) structures of chain 4 of ribosome (PDB code 2i2v, chain 4). The RMSD between the two structures is 3.0Å.

The possibility of using effective connectivities to bias the sampling towards native states opens the way for investigation of folding dynamics using the description of non-specific interactions that we have discussed here. In addition, the fact that the structure profile can discriminate the correct fold from very similar structures, as is necessary in folding, and is predictable to quite good accuracy suggests its incorporation into existing powerful tools of protein structure prediction to exploit the information encoded in the profile. In future studies we will address the problem of reconstruction when the effective connectivity vectors are not known exactly but predicted from the amino acid sequences.

**Acknowledgements**
We gratefully acknowledge financial support by the *Deutsche Akademische Austauschsdienst*, grant number D/08/08872, and *The British Council*, grant number ARC 1319.